FORC+ Analysis of Perpendicular Magnetic Tunnel Junctions

Joseph B. Abugri, [1,2] P. B. Visscher, [1,3] Subhadra Gupta, [1,4] P. J. Chen[5] and R. D. Shull[5]

[1]Center for Materials for Information Technology, University of Alabama, Tuscaloosa, Alabama 35487, USA

[2]Department of Electrical and Computer Engineering, University of Alabama, Tuscaloosa, Alabama 35487, USA

[3]Department of Physics and Astronomy, University of Alabama, Tuscaloosa, Alabama 35487, USA

[4]Department of Metallurgical and Materials Engineering, University of Alabama, Tuscaloosa, Alabama 35487, USA

[5]National Institute of Standards and Technology, Gaithersburg, MD 20899-8552, USA

Abstract

We have studied magnetic tunnel junction (MTJ) thin-film stacks using the First Order Reversal Curve (FORC) method. These have very sharp structures in the FORC distribution, unlike most particulate systems or patterned films. These structures are hard to study using conventional FORC analysis programs that require smoothing, because this washes out the structure. We have used a new analysis program (FORC+) that is designed to distinguish fine-scale structure from noise without the use of smoothing, to identify these structures and gain information about the switching mechanism of the stack.

1. Introduction

The use of the First Order Reversal Curve (FORC) method for characterization of magnetic systems has grown rapidly in the last couple of decades[1-4]. Because it was originally motivated by a model of a magnetic system as a collection of independent particles ("Preisach hysterons") with rectangular hysteresis loops, each completely characterized by a down-switching field $H_R$ and an up-switching field $H$ (Fig. 1), it is not surprising that it is easiest to use for systems such as patterned films that can be reasonably approximated in this way because the magnetostatic and exchange interactions are weak. In such systems, the FORC distribution $\rho$ computed from the FORC curves $M(H_R, H)$ (see Sec. 2 below) as

$$\rho(H_R, H) = -\frac{1}{2}\frac{\partial^2 M(H_R, H)}{\partial H \, \partial H_R} \qquad (1)$$

can be interpreted as the density of hysterons with switching fields $H$ and $H_R$. However, it has become apparent that the FORC distribution gives useful information about unpatterned films[5,6,7] as well, although the features seen in the distribution $\rho$ are different from those of particle assemblies, and cannot be interpreted in the same way. In this paper we argue that the methods that have been developed for the computation and display of the FORC distribution for particle assemblies may not be appropriate for other systems. In particular, noise is a serious problem in FORC analysis because a small amount of noise in the magnetization *M* produces a large amount of noise in its numerical second derivative. Usually this is dealt with by "smoothing", averaging $\rho$ over several field points $(H_R, H)$[8, 9]. This works well when the field dimensions of the structures are large compared to the field increment *ΔH*, *i.e.*, when there is a broad distribution of switching fields. However, in the case of unpatterned films, switching is often quite abrupt (see for example Fig. 3, below), so $\rho(H_R, H)$ has variations on a very small scale, and it is not practical to make $\Delta H$ much smaller than this and perform smoothing. In this paper, we will use a visualization program ("FORC+")[10,11] that makes it possible to distinguish these features from noise without doing any smoothing, hence without washing out the features. This is possible by displaying the raw data in a color scheme in which $\rho = 0$ corresponds to black (or white, in figures designed to be printed on white paper), and positive and negative $\rho$ are represented by complementary colors (*i.e.*, colors that add to white: orange and blue-green, in this paper). Then noise will appear as a finely divided mixture of complementary colors, which appears grey from a distance. In a sense, the human eye automatically does the averaging. A region in which $\rho$ has a positive average will have an orange hue, and one in which $\rho$ is negative will have a blue-green hue.

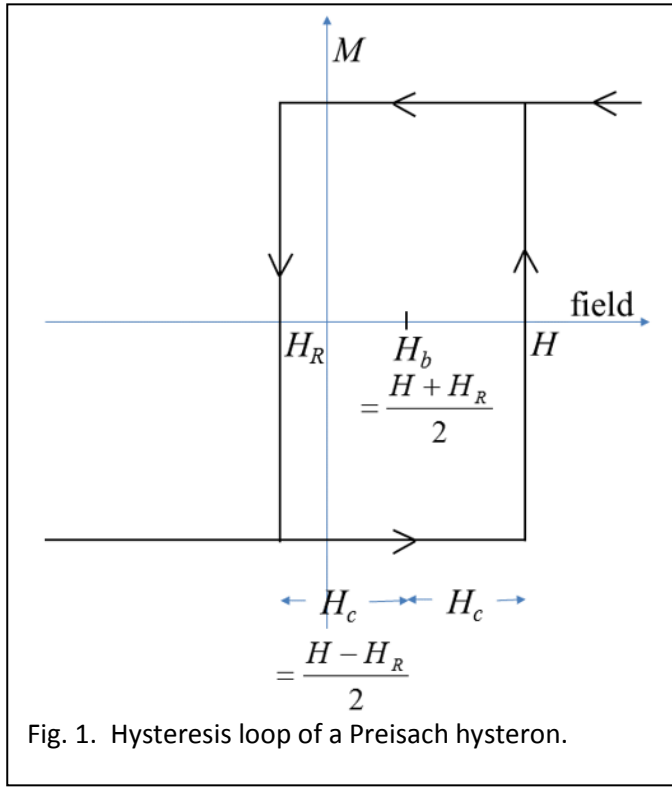

Fig. 1. Hysteresis loop of a Preisach hysteron.

The micromagnetic explanation[5] for the sharpness of the FORC features of unpatterned films involves domain nucleation and the growth of labyrinthine domains[12]; the ability to visualize these sharp features thus enables us to make inferences about the switching behavior without having to do actual real-space imaging.

2. Brief introduction to FORC

The method of First Order Reversal Curves (FORCs) is a powerful and practical tool that has been used for a number of years to extract information from magnetization measurements. It provides information about hysteretic behavior, such as the joint probability distribution of coercivity and bias field that cannot be obtained from the major loop (M-H loop) alone. A FORC is measured (on an alternating gradient magnetometer (AGM) in the present work) by saturating a sample in a large positive applied field $H_{sat}$, decreasing the field to a "reversal field" $H_R$, then sweeping the field slowly back to $H_{sat}$, while recording the magnetization *M(H_R, H)* at discrete values of the field *H*.[1] This process is repeated for many values of $H_R$ yielding a series of FORCs defining the function *M(H_R,H)*. The FORC distribution is defined as the crossed partial derivative of the magnetization with respect to $H_R$ and *H* (Eq. 1). A FORC diagram is then a contour or color-density plot of *ρ(H_R,H)* – we can interpret it as a probability distribution in the coercivity and bias if we transform coordinates to the coercivity $H_c \equiv (H - H_R)/2)$ and the bias field $H_b = (H + H_R)/2$, as indicated in Fig. 1.

The "FORC+" program[10] we use in this paper differs from previous FORC analysis programs in two important ways. First, as mentioned in the Introduction, we can analyze the fine structures that appear in FORC analysis of sheet films, even if they are only a few pixels wide, because the use of a complementary-color scale makes smoothing unnecessary. Second, FORC+ displays the reversible as well as the irreversible properties of the system. The usual FORC distribution includes only the irreversible part of our system – if a system is reversible, *M(H_R, H)* is independent of $H_R$ and therefore its derivative, the Preisach density (Eq. 1), vanishes. We can separate the reversible and irreversible parts by computing the switching field distribution (SFD)

$$SFD(H) \equiv \frac{\partial M(H,H)}{\partial H} = \frac{\partial M(H_R,H)}{\partial H}\bigg)_{H_R=H} + \frac{\partial M(H_R,H)}{H_R}\bigg)_{H_R=H} \quad (2)$$

The second term vanishes in a reversible system, so it is natural[10] to regard it as the irreversible SFD and the first term as the reversible SFD (the R-SFD):

$$R\text{-}SFD(H) \equiv \frac{\partial M(H_R,H)}{\partial H}\bigg)_{H_R=H} \quad (3)$$

The theoretical advantage of this formulation is that one can prove that the FORC+ distribution (by which we mean the Preisach distribution $\rho$ plus the R-SFD plus the saturation magnetization $M(H_R, \infty)$) uniquely determines the original FORC curves. The first derivative (Eq. 3) provides the boundary condition that allows us to integrate the second derivative, and the boundary condition $M(H_R, \infty)$ allows us to integrate the resulting first derivative and recover the original function. FORC+ plots this R-SFD (computed as a simple finite difference) (Fig. 7, below) just above the Preisach distribution, so the irreversible and irreversible parts can be visualized together.

3. Synthesis of multilayer stack

Here we use the FORC method to study a full stack structure for a perpendicular magnetic tunnel junction (pMTJ), shown in Fig. 2.[13] The sample was deposited onto a 20 cm diameter thermally oxidized wafer using a Singulus TIMARIS sputtering system[14] with a base pressure less than $8 \times 10^{-7}$ Pa ($6 \times 10^{-9}$ Torr).

The pMTJ stack was deposited onto a thermally oxidized 300 mm Si substrate with an oxidized thickness of 100 nm. The layer structure is as follows (with the numbers in parentheses being the thickness in nanometers):

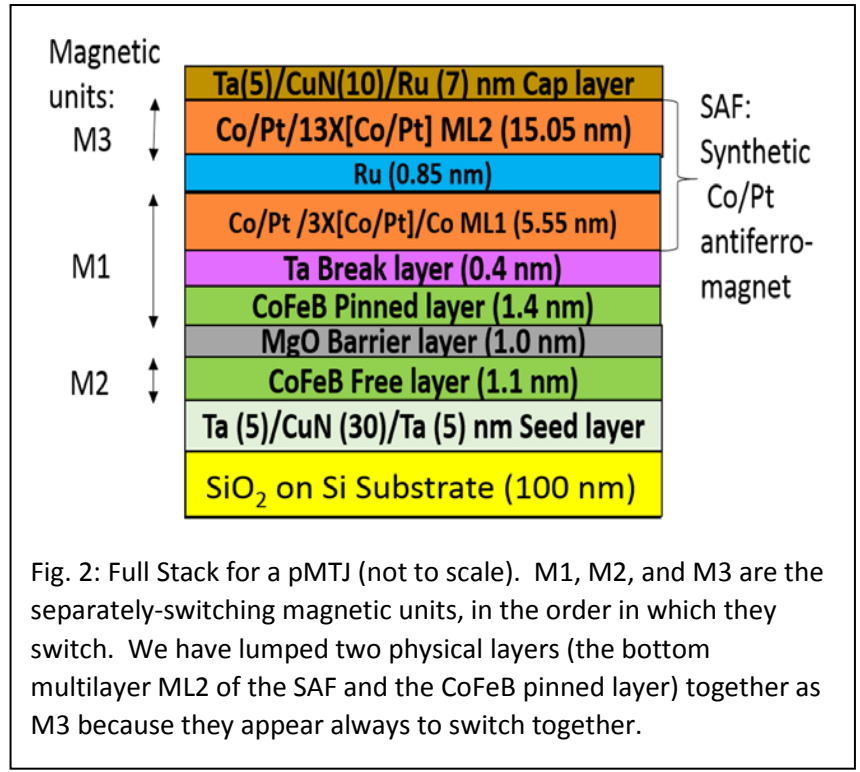

Fig. 2: Full Stack for a pMTJ (not to scale). M1, M2, and M3 are the separately-switching magnetic units, in the order in which they switch. We have lumped two physical layers (the bottom multilayer ML2 of the SAF and the CoFeB pinned layer) together as M3 because they appear always to switch together.

Si/SiO$_2$/Ta (5)/ CuN (30)/ Ta (5)/ Co$_{20}$Fe$_{60}$B$_{20}$ (1.1)/ MgO (1.0)/Co$_{20}$Fe$_{60}$B$_{20}$ (1.4)/ Ta (0.4)\ Co (1) \ Pt (0.8)\ 3x [Co (0.25)\ Pt (0.8)]\ Co (0.6)\ Ru (0.85)\ Co (0.6) \ Pt (0.8)\ 13x [Co (0.25)/ Pt (0.8)]/ Ta (5)/ CuN (10)/Ru (7).

In this full pMTJ stack structure, the lower CoFeB layer is the free layer and is perpendicularly magnetized due to interface anisotropies induced by the MgO barrier layer and Ta seed layer. The pinned or reference layer consists of another CoFeB layer acting as a spin polarizer. It is pinned through a thin Ta layer to a synthetic antiferromagnet (SAF) comprised of a multilayer ML1 (three repeats of (Co/Pt)), which is antiferromagnetically coupled via a Ru layer to ML2, another 13 repeats of (Co/Pt). The sample stack was annealed in a vacuum furnace at a pressure less than 7x10$^{-2}$ Pa (5x10$^{-4}$ Torr) at a temperature of 350 $^o$C for 1 hour in the presence of an out-of-plane magnetic field of 5 kOe.

The TMR (tunneling magneto-resistance) ratio and RA (resistance-area product) of the full pMTJ stack were measured using current-in-plane tunneling (CIPT) in a perpendicular magnetic field. The maximum TMR value was about 80% .[13]

4. Results

We first did a low-resolution FORC scan (field increment $\Delta H$ = 126 Oe) with a large field range (-9 KOe < $H$ < 9 kOe), to see where the interesting features of the Preisach density $\rho$ are located in the $H$-$H_R$ plane. This was done on a Princeton Measurements model 2900 AGM[14], which produces a file with the ($H$, $M$) measurements. This data file can be dragged to the FORC+ program executable[10,11], which displays both the FORC curves and the Preisach density $\rho$ in a window shown (with some additional annotations) in Fig. 3. Fig 3(a) has the FORC curves themselves, as well as the downward branch of the major hysteresis loop (in blue) which connects the beginnings of all the FORC curves. The field region scanned is specified by the user by giving a range of $H_c$ (0 to 5000 Oe in this case) and of $H_b$ (-3000 to 4000 Oe) and the number of FORCs (100, leading to $\Delta H$ = 126 Oe).

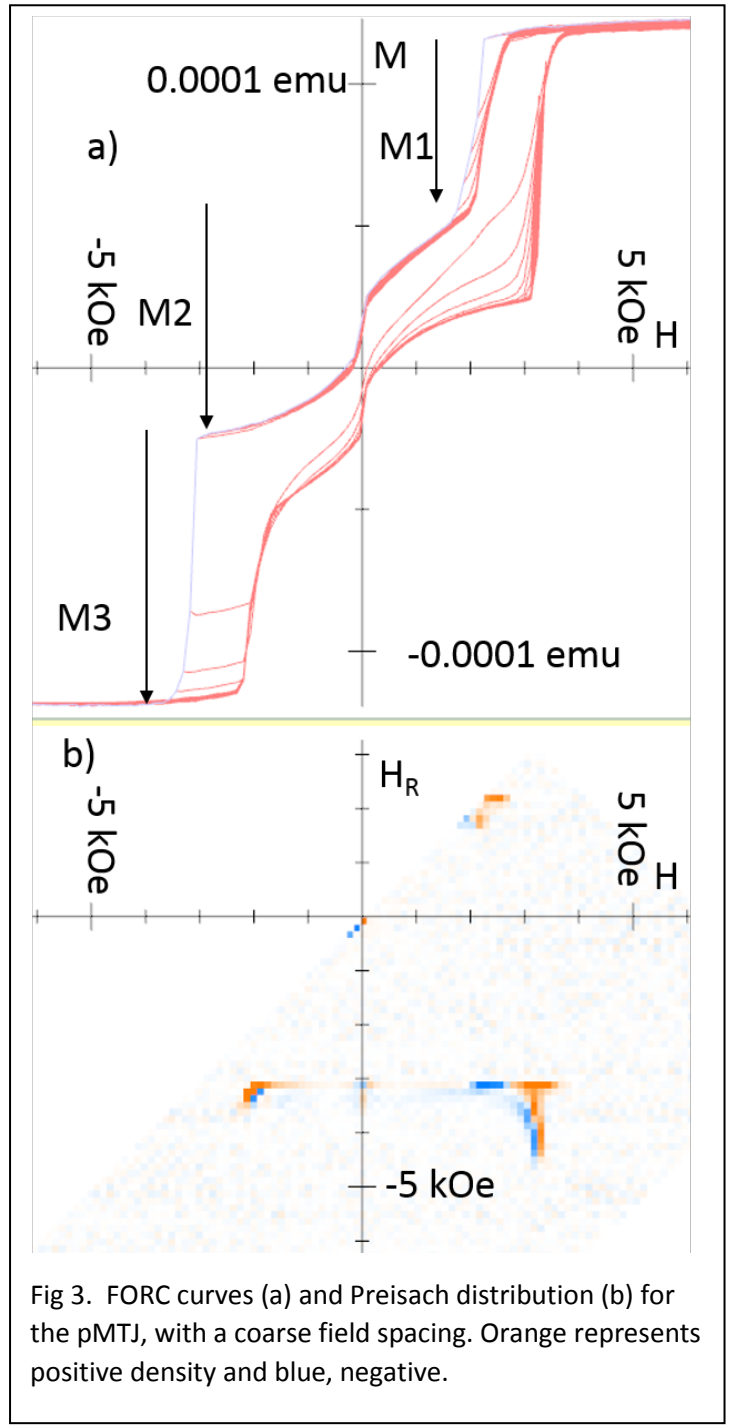

Fig 3. FORC curves (a) and Preisach distribution (b) for the pMTJ, with a coarse field spacing. Orange represents positive density and blue, negative.

The last (lowest) FORC in Fig. 3(a) is essentially the upward branch of the major loop. The blue downward branch can be seen to have three steps M1, M2, and M3. The M1 and M3 steps are fairly sharp, so we identify them with the switching of the strongly coherent Co/Pt multilayers in Fig. 2, the thinner ML1 corresponding to the smaller jump magnitude (and first to switch) M1 (which also includes the pinned CoFeB layer), and the thicker ML2 corresponding to the larger jump magnitude M3. That leaves the free layer, which must produce the more-complicated step M2. To help understand the structures in the Preisach plane (Fig 3(b)), the FORC+ program allows one to move a cursor around the Preisach plane while another cursor shows the corresponding position in the FORC curves. In this way we can identify regions of the Preisach plane in which each magnetic unit is switching, as indicated in Fig 4. The blue arrow pointing downward and to the left follows the major loop, and we have indicated where units M1 and M3 switch down (rather abruptly) and the region where M2 switches down more gradually. As we follow a FORC curve to the right, the layers switch back as indicated by the labeled regions. The switching of M1 and M3 is sudden, so the "Switch M1" and "Switch M3" regions are tall narrow rectangles, whereas the "Switch M2" labels refer to large regions bounded by the other rectangles.

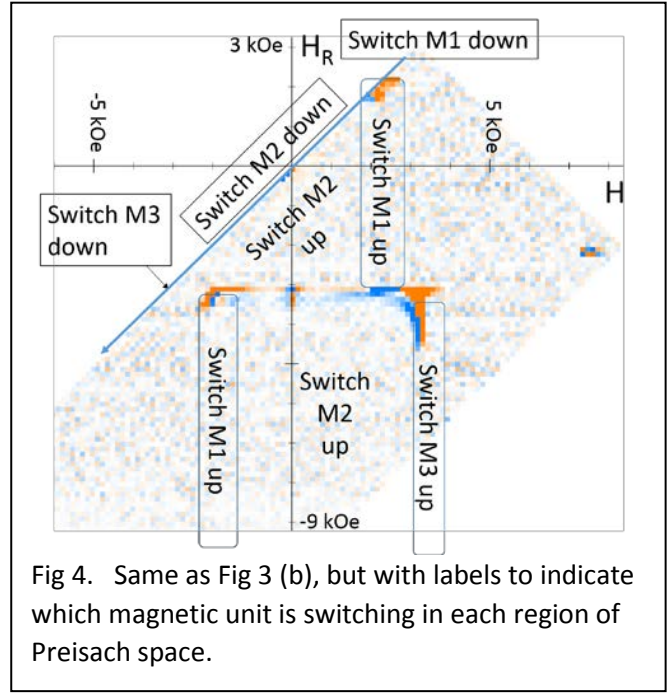

Fig 4. Same as Fig 3 (b), but with labels to indicate which magnetic unit is switching in each region of Preisach space.

The reader will note there are two regions labeled "Switch M1 up" -- in the upper such region, the other units are unswitched (up) while M1 switches up and down, whereas in the lower such region the other layers remain switched down while M1 switches up.

Although the details are clearer in higher-resolution scans, several features stand out in the Preisach distribution (Fig 3(b)). There are several structures we will call "dipolar tails" because they hang down in the $H_R$ direction (constant $H$) and have positive density on one side and negative density on the other. This structure was seen in a single Co/Pt multilayer by Davies et al[5,6,7], who referred to it as a "vertical valley-peak pair". In fact, we can regard the part of our FORC measurement with $H_R$ above about 1 or 2 kOe as essentially the FORC of a single layer M1 – the other layers do not begin to switch in this region, and contribute only a constant bias field. We made a higher-resolution AGM scan in this region (Fig 5) with a much smaller field increment 26 Oe. The effect of the free layer is apparent in the FORC curves (nonzero slope at the lower left) but because the free layer is essentially reversible in this region, this has no effect on the Preisach distribution ρ (Fig 5(b)). The FORC curves are essentially the minor loop corresponding to the switching of this layer. The FORCs strongly resemble those found by Davies et al[5], as does the Preisach distribution. In the insert at the right, we zoom in on the interesting structure in the Preisach distribution (this is done with a single keystroke 'Z' in FORC+).

There is a horizontal positive ridge (labeled "1") and a dipolar tail ("2" and "3"), rather faint in comparison to the others discussed below. Davies et al[5] noted that this tail is very surprising, since it extends in $H_R$ below the apparent lower saturation field of layer ML1, where there should be no dependence on $H_R$ and therefore the derivative ρ should vanish. They offered a micromagnetic explanation for this phenomenon, that there are tiny un-reversed domains ("nuclei") which are too small to be seen in the magnetometry, but facilitate the re-reversal of the ML as $H$ increases. Because they are eliminated only by going to very low $H_R$, this gives the system a hidden $H_R$-dependence, allowing the derivative (the Preisach density ρ) to be nonzero. We will refer to this effect as "holdout bias" because the nuclei that hold out against a strong reversing field shift (bias) the field $H_{sw}$ at which the rest of the ML re-reverses. We have found a way to quantify the holdout bias effect in a very simple model, and show that the resulting Preisach density has the form seen in the experiments.

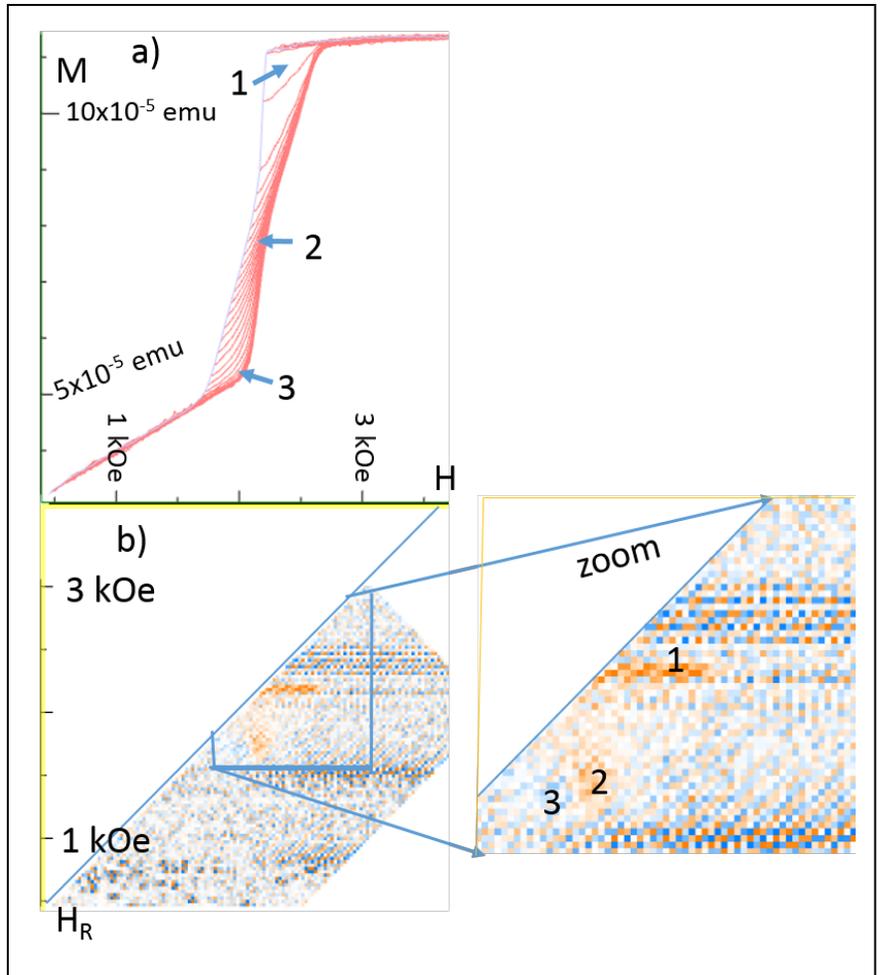

Fig 5. FORC curves (a) and Preisach plots (b) produced by using smaller field ranges (0<$H_c$<1000 Oe, -1500 Oe<$H_b$<3000 Oe), $\Delta H$ = 26 Oe. We have labeled three regions (1, 2, and 3) of the FORC curves, and the corresponding points in the Preisach distribution. Note that there is a high-noise region just above point "1", but it is clearly distinguishable from the positive Preisach density feature at "1" because it has a mixture of positive (orange) and negative (blue) pixels. These figures have a white background for printing – it is actually easier to see the structures on a computer display with a black background[11].

First, we write an equation reflecting the fact that the ML switches up from negative saturation at a field $H_{sw}(H_R)$:

$$M(H_R, H) = -M_s + 2M_s \theta(H - H_{sw}(H_R)) \quad (4)$$

where $\theta$ is the Heaviside step function (zero for negative argument, 1 for positive) whose derivative is the Dirac delta function. The holdout bias function $H_{sw}(H_R)$ could be modeled as $H_{sw}(H_R) = H_{sw}^{-\infty} - H_{HB} \exp\left(\frac{H_R}{h}\right)$, where $H_{sw}^{-\infty}$ is the saturation switching field, $H_{HB}$ is a holdout bias amplitude, and $h$ is a tail length, but we will not assume a specific model here. A possible form is sketched in Fig. 6. To get the Preisach density, we take the derivative

$$\frac{\partial M(H_R, H)}{\partial H_R} = -2M_s \frac{dH_{sw}(H_R)}{dH_R} \delta(H - H_{sw}(H_R)) \quad (5)$$

and another derivative

$$\rho = -\frac{1}{2}\frac{\partial^2 M(H_R, H)}{\partial H \partial H_R}$$
$$= M_s \frac{dH_{sw}(H_R)}{dH_R} \delta'(H - H_{sw}(H_R)) \quad (6)$$

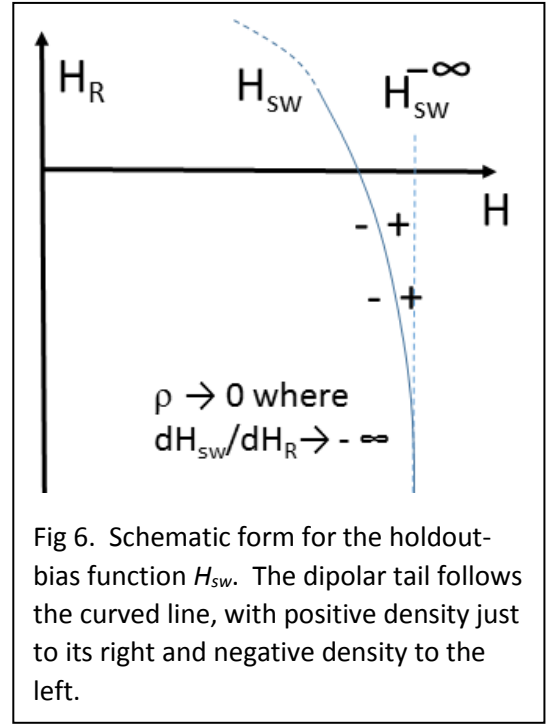

Fig 6. Schematic form for the holdout-bias function $H_{sw}$. The dipolar tail follows the curved line, with positive density just to its right and negative density to the left.

where $\delta'$ is the derivative of the Dirac delta function, i.e. negative on one side and positive on the other side of the curve labeled $H_{sw}$ in Fig. 6, in agreement with the experiments. The factor $dH_{sw}/dH_R$ is normally negative, putting the + signs on the right in Fig 6. This factor makes the amplitude of the dipole tail vanish as $H_R \to -\infty$, also as seen in the experiments. We have dashed the $H_{sw}$ curve above the saturation region to remind the reader that this model needs to be modified there, and we note also that to do a real quantitative fit we would need to spread the $\delta$ function out into something like a Gaussian, so its derivative would be a Hermite function $(\sim x \exp(-x^2))$, but the present calculation gives a quick way to understand the qualitative effect of holdout bias.

Because of the resemblance to the results of Davies *et al*[5], we can conclude with some confidence that the ML1 layer in our sample switches by a similar nucleation-and-domain-wall-motion mechanism, though we are not able to directly image the domains. The horizontal ridge ("1" in Fig 5) is also seen and discussed in ref. 5; we will not try to analyze it further here.

Surprisingly, we find that this form of bipolar tail occurs in three structures in the MTJ Preisach distribution, each associated with the switching of a Co/Pt multilayer (at the top of a "Switch M1" or "Switch M3" rectangle in Fig. 4). The only other notable structures in the Preisach plane are at $H=0$, and are associated with the switching of the free layer (M2). One is at $H_R = -3$kOe where unit

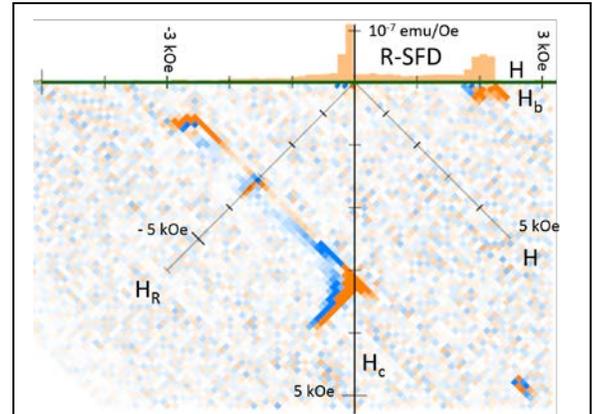

Fig 7. The reversible (switching field distribution, above) and irreversible (Preisach distribution, below) parts of the FORC+ display. The color scales are chosen so that each saturated orange pixel represents exactly the same amount of magnetic moment.

M3 switches down, and the other is at $H_R = 0$ near the zero-coercivity line. These are very faint, if we look only at the FORC distribution. However, recall that this distribution includes only irreversible phenomena – they are faint because the switching of the free layer is mostly reversible. This is apparent from the FORC curves (Fig 3(a)), but FORC+ can plot the reversible switching field distribution (R-SFD) so we can see the reversible effects directly. It is useful to plot the reversible and irreversible distributions together – Fig 7 shows the R-SFD (top) as a bar graph with the same color scheme as the irreversible Preisach distribution below it. This allows us to use the same field axis (horizontal) for the R_SFD as for the $H_b$ axis of the Preisach distribution, if we let the $H_c$ axis point downward. We have normalized the scales so that each pixel of saturated orange color in the R-SFD represents exactly the same amount of magnetization as each saturated pixel in the Preisach distribution, so we can get a sense for the degree of reversibility from the relative amount of color in the two graphs (in this case, the amounts of reversible and irreversible magnetization are about equal). This avoids having to make an artificially sharp conceptual distinction between reversible and low-coercivity irreversible material – when the coercivity becomes zero (and the material disappears from a conventional FORC display), in Fig. 7 it simply moves from just below to just above the horizontal axis. In order to make the R-SFD a little more visible, we have halved the intensity of the orange in it, which doubles the bar height if we keep the relative normalization correct. In order to get a good view of the R-SFD, though, we need to grossly oversaturate the Preisach display so that even the noise is saturated (Fig. 8, where the oversaturation = 64). It is then seen that there are two sharp reversible peaks, one at $H = 0$ from the switching of the free layer, and one at $H = 2$ kOe from the switching of multilayer M1. There is also a broad background, which probably also arises from the free layer. There appears to be no reversible aspect to the switching of the thicker multilayer M3.

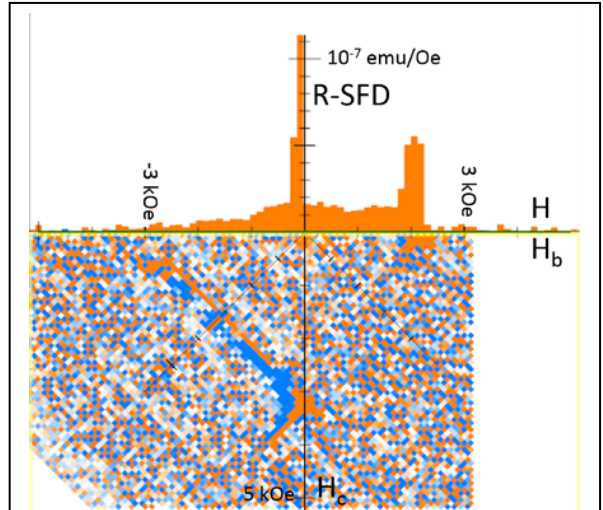

Fig 8. The reversible switching field distribution (R-SFD) on a larger scale (more color per emu) than in Fig. 7. The lower part still shows the irreversible part of the distribution, but is so oversaturated (64x) that it no longer conveys a correct impression of the relative amounts of reversibility and irreversibility.

5. Conclusion

We have applied a new method for FORC analysis that is particularly well suited for the study of unpatterned films, because it does not require smoothing in field and therefore can resolve the very sharp structures due to domain wall motion that are typically seen in the FORC diagrams of unpatterned films. It also allows us to visualize the reversible and irreversible parts of the system together, without making an artificial distinction between reversible and low-coercivity materials.

Acknowledgements

This work was supported in part by the Center for Materials for Information Technology (MINT), and Departments of Electrical and Computer Engineering, Metallurgical and Materials Engineering, and Physics and Astronomy. We thank B. Ocker and S. Tibus (Singulus) for providing the sample.

[14]Tradenames and manufacturers of equipment are given in this paper in order to fully describe the experimental conditions, and does not imply an endorsement by the authors or their organizations.